\documentstyle[aps,prl,epsfig]{revtex}
\title{Prospects for Forbidden-Transition Spectroscopy and Parity Violation Measurements \\
using a Beam of Cold Stable or Radioactive Atoms} 
\author{ S. Sanguinetti, J. Gu\'ena, M. Lintz, Ph Jacquier, A. Wasan and  M-A.~Bouchiat}
\address{\begin{center} 
                Laboratoire Kastler Brossel \footnote{Laboratoire de l'Ecole
Normale Sup\'erieure associ\'e au CNRS (UMR 8552) et \`a l'Universit\'e Pierre et Marie
Curie}, D\'epartement de
                Physique de l'Ecole Normale Sup\'erieure,\\ 
                 24 Rue
               Lhomond, F-75231  Paris  Cedex 05, France \\
(February 28 2003)
                \end{center}}
     
\begin{document}
\maketitle
\begin{abstract}
 Laser cooling and trapping offers the possibility of confining a sample of radioactive atoms in free space. Here, 
we address the
question of how best to take advantage of cold atom properties to perform the observation of as highly forbidden a line
as the 6S-7S Cs transition for achieving, in the longer term, Atomic Parity Violation measurements in radioactive alkali
isotopes. Another point at issue is whether one might do better with stable, cold atoms than with 
thermal atoms. To compensate for the large drawback of the small number of atoms available in a trap, one must take
advantage of their low velocity. To lengthen the time of interaction with the excitation laser, we suggest choosing a geometry
where the laser  beam exciting the transition is colinear to a slow, cold atomic beam, either extracted from a trap or prepared
by Zeeman slowing. We also suggest a new observable physical quantity manifesting APV, which presents several
advantages: specificity, efficiency of detection, possibility of direct calibration by a parity conserving quantity of a similar
nature. It is well adapted to a configuration where the cold atomic beam passes through two regions of transverse, crossed electric
fields, leading both to differential measurements and to strong reduction of the contributions from the 
$M_1$-Stark interference signals, potential sources of systematics in APV measurements. Our evaluation of signal to
noise ratios shows that with available techniques,  measurements of transition amplitudes, important as required tests of
Atomic Theory should be possible in $^{133}$Cs with a statistical precision of $10^{-3}$ and probably also in Fr isotopes for
production rates of $\gtrsim 10^6$ Fr atoms s$^{-1}$. For APV measurements to become realistic, some practical realization of the
collimation of the atomic beam  as well as multiple passages of the excitation beam matching the atomic beam looks essential.                  

\end{abstract}
\section{Introduction : Motivations} 
Atomic Parity Violation (APV) measurements have proved successful for
probing at low energy one of the most fundamental predictions of the Standard Model
(SM), namely the existence of a weak electron-nucleus interaction mediated by the exchange of neutral gauge bosons
$Z_0$ \cite{bou97,woo97,ben99}. Up to now the efforts have been focused on the comparison between the experimental
determination of the weak charge of the atomic nucleus, $Q_W$, and its SM prediction at the 0.5$\%$ level of precision, the
cesium atom lending itself to the most precise comparison 
\cite{blu90,der01,mil01}. Actually, it looks somewhat too early to assert definitely  either the absence or existence of a
deviation, most likely less than 2.5
$\sigma$ \cite{mil01,fla02}. On the other hand, it has not been possible, yet, to test another important SM prediction
concerning the variation of 
$Q_W$ along a string of isotopes belonging to the same element. An original experimental approach is currently
pursued for rare-earth elements namely Yb \cite{dem95} and Dy\cite{bud99}, but it also would be extremely valuable 
to extend the measurements which have proved successful for natural cesium,
$^{133}$Cs (the sole stable Cs isotope), to a few of its numerous radioactive isotopes, as well as to other alkali
isotopes, more excitingly radioactive francium. With Z=87, francium is
expected to lead, due to the fast increase with Z \cite{bou97}, to APV effects 18 times larger than cesium, while it does not look
unrealistic to have a theoretical prediction of its weak charge as precise as that for cesium
\cite{fla95}. Indeed, atomic structure calculations for alkali are (barring H and He) the most precise available. This, added to
the fact that many isotopes can be produced, makes this element often considered as one of the most interesting candidates for
forthcoming experiments.  Moreover, since up to now, the nuclear anapole moment \cite{fla85} has been detected only for $^{133}$Cs
\cite{woo97} (an even neutron-number isotope), it is important to measure it for another isotope (preferably one with an odd
neutron-number). Regardless of APV, measurements on the forbidden line in alkali-metal atoms are
important since forbidden magnetic dipole amplitudes are ``the most sensitive among electromagnetic transition amplitudes
to the accuracy of the relativistic description of an atomic system'' \cite{joh99}, i.e. rigorous tests of atomic theory.    

 Francium, and more generally short-lived radioactive atoms, either obtained from a radioactive source or produced on line by an
accelerated ion beam colliding a target, are produced at a limited rate, with a thermal, or even  superthermal, velocity
distribution. In order to perform APV measurements the first prerequisite is to avoid their spreading out in space and their
loss inside the wall. Only the radiative cooling and trapping techniques \cite{met} possibly combined with Light Induced Atomic
Desorption (LIAD) \cite{moi99} can succeed in this kind of operation. Several successful attempts to load radioactive alkali
atoms in a neutral atom trap have been reported with $^{21}$Na \cite{ztlu94}, $^{38}$K$^m$, $^{37}$K \cite{beh97},
$^{79}$Rb \cite{gwi94}, $^{135}$Cs \cite{vie02}, $^{207-211}$Fr \cite{sim96}, $^{221}$Fr \cite{ztlu97}.  Observation
of several allowed Fr transitions  has 
been realized for atoms trapped inside a MOT, leading to precise spectroscopic measurements
\cite{oro00}. But never, yet, has it been reported for a transition as highly forbidden as the Fr 7S-8S transition. Therefore,
before attempting APV measurements with cold atoms, a preliminary - and by no means straightforward - objective consists
in observing the 6S-7S transition with trapped Cs atoms. Since the precise value of the parity conserving transition amplitudes, in
particular the Stark induced amplitude associated with the vector polarizability $\beta$, is still a somewhat open question
(see below), as an assessment of the potential of trapped atoms for this kind of experiment, we suggest a new
precision measurement of the ratio
$M_1^{hf}/\beta$.  Here, the magnetic dipole amplitude $M_1^{hf}$ induced by hyperfine interaction serves as a precisely
known amplitude used for calibration
\cite{bou881,joh01}. This would be all the more precious since the previous measurements 
\cite{ben99} were made in acrobatic conditions (background equal to 100 times the signal \cite{ben99}) and have led to a
result for
$\beta$  which differs from a recent independent semi-empirical determination by (0.7 $\pm$ 0.4)$\%$ \cite{vas02}.
Though small, such a difference is sufficient to narrow the gap between theoretical and experimental values of 
$Q_W({\rm Cs})$ from 2.2 to 0.9$\sigma$. A measurement of $M_1^{hf}/\beta$ in cesium will allow us to assess the
feasibility of similar measurements in francium, knowing the production rate. Finally, we also attempt to evaluate the
feasibility of an even more ambitious project, namely a new high precision measurement of the parity violating electric
dipole amplitude $E_1^{pv}$ in cesium, and hence 
$Q_W({\rm Cs})$ by an independent method using cold atoms. Indeed, such an independent measurement would be
extremely valuable as a cross check of this fundamental quantity
\cite{cas99}. Beyond this, we cannot understate how welcome a measurement of $Q_W({\rm Fr})$ would be, if some day
feasible.

Our paper is a prospective work suggesting preparatory experiments for much more ambitious projects. Once a sample of
cold alkali atoms is produced at the center of a trap, there remains a still unsolved point at issue~: 
 what is the best way to use it for exciting and probing the forbidden Cs 6S-7S transition, or the analogous 7S-8S
transition  in Fr, in the Stark electric field necessary to previous APV measurements? Even for the stable
isotope $^{133}$Cs, the biggest difficulty is linked to the small number of atoms
available in a trap. It is the purpose of the present paper to quantify such a difficulty by making
comparisons with conditions realized in previous APV experiments performed with stable, thermal atoms and to suggest
an experimental approach using their different specific properties. We suggest advantageous means to exploit their low
velocity and we also propose a new physical observable which, we believe, is well adapted to this situation. It is
shown to be well suited to the measurement of first $M_1^{hf}/\beta$ and later $E_1^{pv}/M_1^{hf}$. Concerning the
measurement accuracy, in an approach of this kind two parameters play an essential role~: i) the number of atoms present at
a given time in the interaction region, ii) the probability for such an atom to contribute to the APV signal, which takes into
account the nature of the physical observable, and both the excitation and detection efficiencies. We have not
found the ideal compromise between simplicity and outstanding performances, transferable from Cs to Fr. According to
the exact goal to be reached the experimental scheme to be chosen will probably have to change.  We consider three different
experimental approaches, all of them relying on the production of a cold, slow atomic beam. They differ by the method
of production of the beam and its parameters. This will appear explicitly in \S  II. The observable physical quantity
is presented in \S III, while \S IV describes a  method to suppress the
dangerous systematic effect which might arise from the Stark-$M_1$ interference effect when one wants to measure
$E_1^{pv}$. Finally  (\S V), we make predictions for the Signal to Noise ratio for measuring the interesting physical quantities
mentioned above, in these three different, well-defined and realistic, experimental configurations. 
 
\section{Use of a slow and cold atomic beam excited by a colinear laser beam}
The energy levels and wavelengths relevant to APV measurements and to laser trapping operation for both cesium 
and francium are shown on Fig. 1. A precise value of the  measured energy difference between the Fr 
8S$_{1/2}$ and 7S$_{1/2}$ levels is given in \cite{sim99}. 
Performing the PV measurements {\it inside} an optical molasses or a Magneto Optical Trap (MOT), precisely where
the atoms are cooled and stored presents some inconvenience. Indeed, both laser cooling and APV measurements require
specific conditions which look difficult to reconcile~: for instance, the presence of excited atomic species in the interaction
region and  the interception of the wide cooling light beams by the necessary electrodes are both difficult to avoid. 
Among a large variety of techniques available for
manipulating atomic velocities by electromagnetic forces~\cite{cct98}, several of them offer the possibility of producing a
well collimated beam of  slow, cold atoms. In order to concentrate the discussion on a well defined situation, we shall
choose the example of a pyramidal MOT built according to the judicious and simple suggestion of Lee et al. \cite{lee96},
now used by several groups
\cite{foo98,ari01}. The trap is built with four mirrors, standing as the four faces of an inverted pyramid. A single
cooling beam, circularly polarized,  is enough to create, after reflexion on the four mirrors, the same field configuration as a
standard six-beam MOT. When the pyramidal trap is vapor-loaded, a beam of slow and cold atoms escapes continuously
through the hole bored through the pyramid apex because of local imbalance of intensities. In typical conditions this kind of
device can provide a continuous flux
$\Phi_{at}$ of
$\gtrsim 2\times 10^9$ cesium atoms/s with a mean velocity $ v$ tunable around 10 m/s and a velocity
spread less than 1.5 m/s \cite{ari01}. 
 The transverse velocity spread of the beam is found to lie close to the Doppler transverse limit for cesium $\sqrt {\hbar
\Gamma/m}=0.13$ m/s. 

\begin{figure}
\centerline{\epsfxsize=150mm  \epsfbox{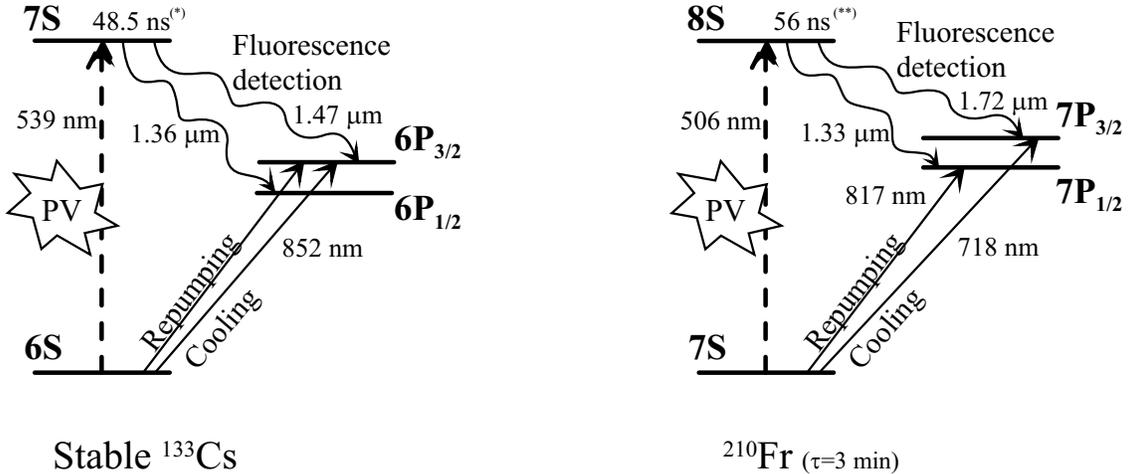}}
\caption{Energy levels and wavelengths relevant for APV experiments and atomic cooling and trapping in cesium and
francium. $(^\star)$ Cs 7S state lifetime from experiment \protect\cite{bou84}, and $(^{\star \star})$ Fr 8S state  lifetime from
theory \protect\cite{bie98}.}
\end{figure}

The pyramidal trap has advantages of low cost and simplicity, but still better performances can be
achieved with more sophisticated devices. In particular, ref.\cite{rii00,dgo02} describes how an even slower and colder
rubidium beam can be obtained using a vapor loaded laser trap which ensures two-dimensional magneto-optical trapping, as well as
longitudinal cooling by a moving molasses (MM-MOT). The average velocity can be as low as  1m/s and the velocity
distribution has been evaluated: $\Delta v / v < 1/10$.     
When either one of those slow atomic beams is illuminated by a co- or counter-propagating narrow line-width cw laser
beam, {\it the dispersion of the longitudinal velocities is small enough for all atoms excited on the 6S-7S transition to belong
to a single velocity class}. Moreover, at optimum alignment, the divergence of the atomic beam (26 mrd FWHM) going out of
the pyramidal trap is small enough for avoiding any significant atom loss out of a 1~mm radius laser beam over an
interaction length of 4 cm. In contrast, the larger divergence of the ultra cold beam \cite{dgo02} unavoidably  complicates the
design of the experiment (see proposal 2, \S V B). Other 2D-MOT, among those delivering larger atomic fluxes
\cite{wal98,int96,pfa02}, have, for the present application, the drawback of either a larger divergence or a larger velocity
spread.  On the other hand, the features needed here, high flux, moderate velocity and low divergence are met
by other techniques, namely Zeeman slowing. In particular, the Zeeman-slower apparatus described in ref 
\cite{lis99} has very attractive features~: flux of Cs atoms exceeding 10$^{10}$ s$^{-1}$, very small spread of longitudinal velocities,
$\sim$  1 m/s and much better collimation (divergence angle
less than 1 mrd). This gives the possibility of lengthening the interaction region (see proposal 3 in Table~I and \S V C). 
 Although some adaptation will be necessary to make the system work with radioactive isotopes, depending on their method
of production, it is interesting to evaluate its potentiality as compared to that of a thermal atomic beam or a
vapor cell.

After a short path beyond the trap exit (or the collimation module \cite{lis99}), the atomic beam enters the interaction
region which includes capacitor plates generating the Stark field, with the plane electrodes parallel to the beam direction. 
The first relevant parameter to be compared here to previous experimental configurations is the number of atoms
$N_{at}$ interacting with the excitation laser beam in the interaction region of length $l$.

For an atomic beam experiment  
$N_{at}=\Phi_{at} \times l/v$.  For a vapor at thermal equilibrium (Paris experiment \cite{bou02})
we take into account that only a fraction of the atoms can absorb the resonant light beam which has a spectral width much
smaller than the Doppler width. By averaging the  velocity-dependent transition probability over the
thermal distribution, one finds that  this can be accounted for by the reduction factor~: $R= \sqrt{2 \pi} \Gamma/ 4 \Gamma_D$
(see ref.
\cite{bou75}),  where $\Gamma_D = \omega_0 \sqrt{\frac{kT}{Mc^2}}$, and $\omega_0/2 \pi $ is the transition frequency;
$\Gamma$ is the radiative line width of the 7S state, including both the emission rate of spontaneous and stimulated photons.
In the conditions realized experimentally, we arrive at $R \approx  0.035$.  The factor R yields the
fraction of atoms sufficiently slow to be in interaction with the resonant excitation laser, thus we obtain 
 $N_{Cs}=n_{Cs} V\times R$, $n_{Cs}$ being the cesium vapor density, and V the interaction volume.  

Table~I collects the value
of this important parameter $N_{at}$ expected in the present proposals, for comparison with those obtained in the
experiment having previously yielded APV data in Cs.  It clearly appears that
the effect of the much larger atomic flux available with the thermal beam used by the Boulder group
\cite{woo97,woo99} is counterbalanced  by the much shorter interaction time resulting from a $\sim$ 30 times larger
velocity and an interaction length $\sim$ 50 times shorter to ensure transverse excitation of  the
beam. By comparison, the vapor experiment developed in Paris takes complete advantage of having at one's
disposal a number of atoms in the interaction region up to tens thousands times larger. The thermal beam
experiment compensates for this deficit by use of a huge laser power in the interaction region owing to a Fabry-Perot cavity
with a finesse of $\sim 10^5$. From the point of view of systematics each approach has its advantages and its drawbacks.
 
Cold atomic beams of several kinds have been described in the literature. Since our purpose is to assess how well each one is  
adapted to performing APV measurements, with comparison in view, we introduce    
a quality factor aiming at taking into account the divergence of the atomic beam, $\frac{\Delta v_{\perp}}{v },$ the main limitation to 
measurement efficiency. We first define the optimum length of interaction $l_{opt}$, as the length over which the atomic beam radius
$r$, does not exceed 1mm, a reasonable value for a laser beam radius \footnote{More precisely, $l_{opt}$ is defined by the
condition:
 $  r =  r_0 + \frac{\Delta v_{\perp}}{v }l_{opt} = 1 \; {\rm mm},$  where  
 $r_0$ is the atomic beam radius at the pyramidal MOT or collimator output and $\Delta v_{\perp} =\sqrt{kT_{\perp}/m}  
$.}. For  proposals 1, 2 and 3, we obtain 
$l_{opt} = 4$ cm, 1.5 cm  and 60 cm respectively \footnote{The beam can be horizontal~: the vertical
displacement  $g l^2/ 2 v^2$ over these distances, due to gravity, does not exceed 0.5 mm. }. Then, the quality factor is
defined as the number of atoms in the interaction region of length $l_{opt}$, namely $f_{APV}= \Phi_{at} \times  l_{opt} /v $.

\begin{table}
\caption{Number of atoms in the interaction region for the cold atomic beam proposals compared to the
previous situations in which APV measurements have been performed. Figures collected in the last column,
clearly illustrates that, for stable atoms, a vapor experiment presents from the outset a large advantage. }


~~~~~~

\hspace{10mm}~~~~~~~~~~~~~~~~~~~~~Atom Flux
~~~~~~~~~~~Velocity~~~~~~~~~~~~~~~~Length~~~~~~~~~~~~~Int. time~~~~~~~~~~~~~~ Number of atoms 

~\hspace{140mm} in the interaction region

\hspace{10mm}~~~~~~~~~~~~~~ ~~~~~~$\Phi_{at}$(at s$^{-1}$)~~~~~~~~~~~~$v$
(cm s$^{-1}$)~~~~~~~~~~~ $l$ (cm)~~~~~~~~~~~~~~~~$\tau(s)$~~~~~~~~~~~~~~~~~~~~~N$_{at}$

~~~~~~~~

~~~~~~~~

\noindent Slow, cold beam  \cite{ari01}~~~~~$2\times
10^9$~~~~~~~~~~~~~$~0.8\times10^3$~~~~~~~~~~~~~~~~~~~4~~~~~~~
~~~~~~~~~~~~$5\times 10^{-3}$~~~~~~~~~~~~~~~$1\times 10^7$

\noindent (Proposal 1)

~~~~~~

\noindent  Ultra cold beam~~\cite{dgo02}~~~~~~$2 \times
10^9$~~~~~~~~~~~~~~~~~~~~ $1 \times 10^2$~~~~~~~~~~~~~~~~~~~~~~~  5~~~~~~~
~~~~~~~~~~~~~~~~~$5 \times 10^{-2}$~~~~~~~~~~~~~~~~~~~$1 \times 10^8 \left(\times \frac{1}{10}
\right)^{\star}$
 
\noindent (Proposal 2)

~~~~~~

\noindent Zeeman-slower \cite{lis99} ~~~~$2.6 \times
10^{10}$~~~~~~~~~~~~~ $9 \times10^3$~~~~~~~~~~~~~~~~~~60~~~~~~~
~~~~~~~~~$6.7 \times 10^{-3}$~~~~~~~~~~~~~~$1.7 \times 10^8$

\noindent (Proposal 3)

~~~~~~

\noindent Thermal beam\hspace{10mm}~~~$~1 \times10^{13}$~~~~~~~~~~~~~~~ $3
\times10^4$~~~~~~~~~~~~~~~~~0.08~~~~~~~~~~~~~~~$2.6 \times 10^{-6}$~~~~~~~~~~~~~~$2.6 \times
10^7$

\noindent (Boulder \cite{woo97})

\noindent completed expt.

~~~~~~~~~~

\noindent
Vapor \hspace{10mm} ~~~~~~~  Density(cm$^{-3})$~~~~~~~~~~Volume(cm$^3$)~~~~~~~~~~
~~~$ R$

(Paris \cite{bou02})~~~~

current expt.
 \hspace{16mm}$2 \times 10^{14}$~~~~~~~~~~~~~~~~~~0.1~~~~~~~~~~~~~~~~~~3.5 $\times
10^{-2}$~~~~~~~~~~~~~~~~~~~~~~~~~~~~~~~~~~~~~~$~7 \times 10^{11}$

~~~~~~~~~

$^{\star}$ {\footnotesize This
additional factor is a rough estimate of the loss occasioned by spreading of the beam, unless special design of the
experiment (e.g. multiple passages of the excitation beam) solves this difficulty.}

\end{table}

\begin{figure}
\centerline{\epsfxsize=100mm  \epsfbox{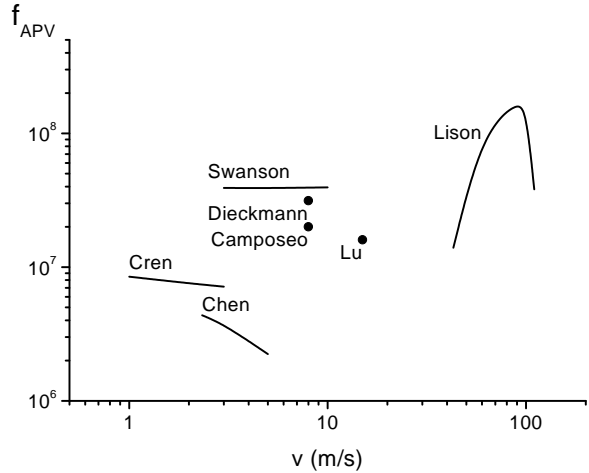}}
\caption{The quality factor $f_{APV}$ versus atomic longitudinal velocity for several alkali sources of cold atomic beams. Camposeo
{\it et al.} \protect \cite{ari01} (Cs), Chen {\it et al.} \protect\cite{rii00} (Rb), Cren {\it et al.} \protect\cite{dgo02} (Rb), Dieckmann
{\it et al.}
\protect\cite{wal98} (Rb), Swanson {\it et al.} \protect\cite{int96} (Rb), Lison {\it et al. }\protect\cite{lis99} (Cs),   
  Lu {\it et al.} \protect\cite{wie96} (Rb). }
\end{figure}
On Figure 2 we plot the quality factor versus the velocity for several beams
of cold  stable atoms chosen among those having a small spread of longitudinal velocities.  The existing designs present themselves
as grouped into three categories~: the ultra-cold beams using a moving molasses \cite{rii00,dgo02}, the cold beams extracted from
a 2D-MOT \cite{ari01,wal98,int96,wie96} and the Zeeman-slowed device using a collimator \cite{lis99}. In view of optimizing 
APV measurements on stable atoms, this last device is expected to lead to the best results, although the pyramidal trap
remains of real interest due to its simplicity and probably better adaptability to radioactive isotopes.
As we noted previously, the performances expected with the cold atomic beams are limited essentially by their divergence.
However one may imagine two means of palliating this kind of difficulty.

i) {\it Multiple passages of the excitation beam~:} it looks possible to widen the interaction region, at fixed density of excitation
energy, by performing forward-backward passages of the beam between two spherical mirrors.  The two mirrors should be pierced,
one for  providing the passage of the atomic beam at the output of the MOT and the other the passage of the
counterpropagating excitation laser \cite{mul}.         

ii){\it   Insertion of a collimator at the output of the MOT~:} It would seem very interesting to insert at the output of a two
dimensional MOT a collimator similar to that described in \cite{lis99}. Besides the beam collimation this device
has the attractive feature of deflecting the atomic beam by a small angle, thus making possible to place the interaction region inside a
Fabry-Perot cavity which provides enhancement of the excitation energy density. 
However we must be aware that a transverse temperature at the output of the collimator less than 50 $\mu$K looks difficult to
achieve. Therefore the divergence of the slow beam remains well above that of the faster Zeeman-slowed beam.

\section{A well adapted observable physical quantity and two interaction regions  }
 The choice of the observable physical quantity which manifests APV also plays an important role, since it determines
the specific nature of the signal (absence or presence of a background), its signature and it also conditions the detection
efficiency. In our first experiment in Paris \cite{bou82}, as well as in our current second-generation one \cite{bou02}, we
have chosen to detect an angular momentum anisotropy in the excited state (either an atomic orientation in the first version,
or an atomic alignment in the latter) providing 
a very specific signal without background. However,  fluorescence detection efficiency of the 7S state orientation was low ($\sim
10^{-3}$), due to the need of polarization analysis on a single fine structure line.  Alignment detection can be conducted efficiently
using stimulated emission detection \cite{bou02,bou96}.  However, in view of the very small number of atoms available in a trap, there
is no possibility of signal amplification by the stimulated emission process advantageously used in a dense vapor. Therefore, with a
cold beam there is a strong incentive for detecting the PV effect on the absorption rate.  

We suggest to create a spin polarization $\vec P_e$ of the atomic beam at the output of the trap in a direction perpendicular
to its velocity. Then, a  specially well adapted observable physical quantity is a contribution to the absorption rate involving
this spin polarization. It results from an interference between the parity-violating electric dipole amplitude $E_1^{pv}$ and
the Stark amplitude induced by a transverse electric field. More precisely, the manifestation of APV would 
\begin{figure}
\centerline{\epsfxsize=120mm  \epsfbox{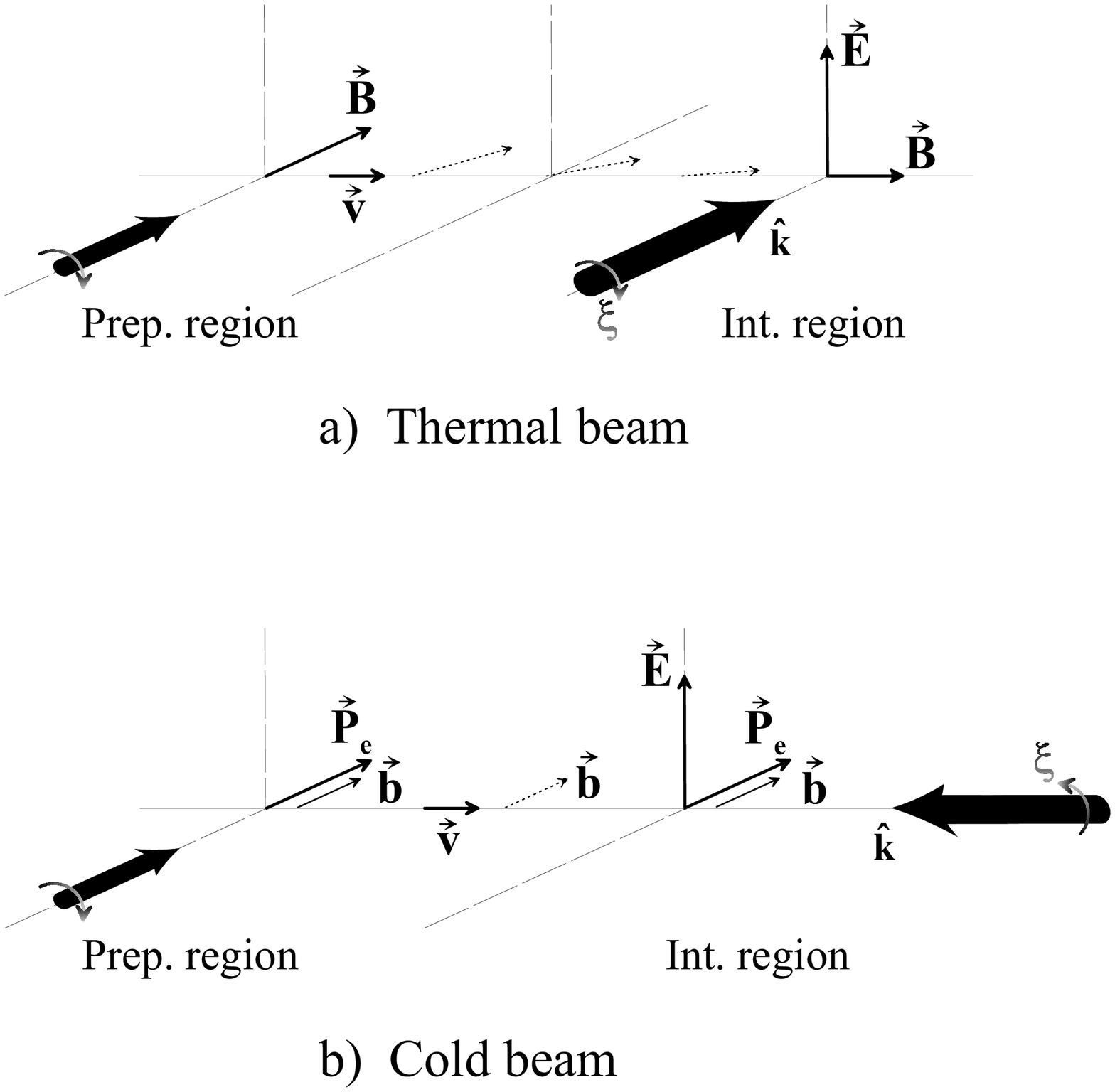}}
\caption{ Schemes of the geometrical configurations relative a) to the Boulder experiment performed with a thermal atomic
 beam \protect\cite{woo97} and b) to the present proposal using a cold and slow atomic beam. Both exploit the spin
polarization
$ \vec P_e$ of the atoms performed in a preparation region and use a transverse Stark
electric field $\vec E$ in the interaction region. Both make use of a circularly polarized excitation beam (helicity $\xi$). 
With the thermal beam, the excitation beam has to be transverse to the velocity and the magnetic field $\vec B$, large
enough to resolve the Zeeman components, has to rotate its direction by $\pi/2$ between the preparation and interaction
regions, while with the cold beam the excitation can be longitudinal and only a small magnetic field $\vec b$, of
uniform direction, is required, to preserve the spin polarization between the two regions. The pseudoscalar
$\vec E \cdot \xi \hat k
\wedge \vec B$ manifesting APV in case a) is replaced by $\vec E \cdot \xi \hat k \wedge \vec P_e$ in case b).     }
\end{figure}
\noindent then rely on the
presence in the absorption rate of the pseudoscalar quantity $ \vec E \wedge \xi
\hat k \cdot \vec P_e$, where $\xi \hat k$ represents the angular momentum of the light beam which excites the transition and
$\vec E$ is the applied static electric field.
 It has the advantage of appearing in the total 
population of the excited
state. It can be detected by monitoring the total intensity of the fluorescence light emitted during the two-step
desexcitation process, involving either the
$6P_{1/2}$ or the $6P_{3/2}$ state. No polarization analysis nor even light filtering (except for stray light) is 
necessary in principle. The APV signal is odd under the separate reversals of the electric field, the spin polarization and the
helicity of the photons which excite the transition.  We relegate to the appendix the derivation of the signal expression in the
most general conditions. Here we present the result in the particular case of $^{133}$Cs (I=7/2), for the experimental
configuration shown in Fig. 4,    
 supposing no magnetic field and a total circular polarization of the excitation beam, $\xi= 2 \,{\rm Im}
\lbrace \epsilon_x^* \epsilon_y \rbrace = \pm 1$ (hence $\vert \hat \epsilon\cdot \hat u \,\vert^2= 1/2 $ whatever $\hat
u \perp \hat k$ ). 
\begin{eqnarray}
& N_{7S} \propto \beta^2 E^2 - \frac{3}{4}  \; ( M'_1  +  \xi \; {\rm Im} E_1^{pv}  )
\; \beta \vec E \wedge \hat k \cdot \vec  P_e  \hspace{10mm} {\rm for \;the}\;\; 6S, F=3
\rightarrow  7S, F=4 \;\;  {\rm line} \;,\\
 & N_{7S} \propto  \beta^2 E^2 - \frac{5}{4}  \; ( M'_1  +  \xi \; {\rm Im} E_1^{pv}  )
\; \beta \vec E \wedge \hat k \cdot \vec  P_e  \hspace{10mm} {\rm for \;the}\;\; 6S, F=4
\rightarrow  7S, F=3 \;\;  {\rm line} \;.    
\end{eqnarray}
$\beta$ denotes the vector polarizability of the transition\footnote{From
the radial matrix elements and the experimental energies compiled in
\cite{dzu01}, we have obtained estimates of the
scalar and vector Fr 7S-8S transition polarizabilities, $\alpha=-361$ ea$_0^3\, ,\beta = 50$ ea$_0^3$ hence
$\alpha / \beta \approx -7.2$ (instead of -261  ea$_0^3$,  27  ea$_0^3$ and  -10  for the Cs 6S-7S
transition).   } and
$M'_1$, the magnetic dipole amplitude, which is the sum of the many-body contribution $M_1$ and that induced by the
hyperfine interaction
$M_1^{hf}$. Here we assume the applied electric field large enough so that the field-independent contribution proportional
to $M_1^{\,'2}$ can be neglected.

 We note that {\it this circular dichroism of a transversally polarized sample}, $\vec E \cdot \xi \hat k \wedge \vec
P_e$, could not be envisaged in a dense vapor where the spin polarization is rapidly destroyed by collisions. By contrast,
(co-)counter-propagation of the atomic and light beams provides the attractive possibility of having both beams passing
{\it through two interaction regions leading to circular dichroism of opposite sign}. For instance, one can choose two
orthogonal directions of $\vec E$ in these two regions, with the direction of $\vec P_e$ taken at $\pm 45^{\circ}$ to the
direction of 
$\vec E$ in one and the other region (See configurations 1 and 2, or 3 and 4, represented in Fig. 4). Then, the difference of
fluorescence rates in those two regions can selectively provide the $\vec P_e $-dependent contribution of interest. In the next
section, we shall show that such a differential measurement also offers the important  additional  advantage of suppressing some
dangerous systematic effects. 

It is important to notice that real time calibration of the PV signal is easy to obtain. By
selecting in the fluorescence rate the contribution odd under the separate reversals of the electric field and the spin polarization, but
even in the reversal of the light helicity, we can isolate the $M'_1$-Stark interference signal. Thereby the amplitude ${\rm Im}
E_1^{pv}$ is directly calibrated\footnote{This calibration procedure performed in each region independently, eliminates the
magnitude of the spin polarization and the exact value of $\vert \vec E \cdot \xi \hat k \wedge \vec P_e\vert$ as well as
other  geometrical parameters (beam spreading, detection efficiency, etc...) which may differ from one to the other region.} in
terms of $M'_1$. If one reminds that
$M'_1 = M_1
\pm M_1^{hf}$,  depending on the hyperfine transition $\Delta F = \pm 1$, we see that absolute calibration of
${\rm Im} E_1^{pv}$ in terms of the theoretically well known amplitude $M_1^{hf}$ is possible. 

 Another observable physical quantity appearing in the absorption rate has been proposed in \cite{bou79}.  It does not require any
spin polarization of the ground state, but it involves the application of a magnetic field
$\vec B$, transverse to the light beam, which enters explicitly into the definition of the pseudoscalar manifesting parity
violation, $\vec E \cdot \xi \hat k \wedge \vec B$. However, for observing this effect the field has to be large enough for
the Zeeman components to be resolved, otherwise compensations occur \cite{poi79}. This is, actually, the APV effect which
has been detected by the Boulder group
\cite{gil86}. In the most recent version of their experiment \cite{woo97} (see Fig. 3-a for a schematic representation of the
configuration), the atomic beam is spin polarized in a preparation zone before entering the interaction region, but a magnetic
field (6.4~G), whose purpose is to resolve the Zeeman lines is still applied, although the atomic spin polarization prepared in
the ground state makes this unnecessary, as Eq. 1 shows. With the same set-up, a much weaker field would be 
sufficient for  preserving the direction of the atomic orientation between the preparation and the interaction regions. This 
would avoid slight line overlap of adjacent Zeeman lines and the associated difficulties.

We, now, want to comment about the conditions to be fulfilled by the magnetic field, which obviously cannot be perfectly
cancelled. There are strict requirements: the magnetic field of the MOT has to be screened. Instead, a
small $\vec B$ field along the direction wanted for the spin polarization 
 is needed in the optical pumping region as well as in the two interaction regions and, consequently, in between those two 
regions: otherwise the rapid spin precession might result in spin disorientation (when the spins do not follow adiabatically
the field direction). Finally, the exact direction of
$\vec P_e$ inside the interaction regions, involved in the pseudoscalar manifesting APV, is actually determined by the
$\vec b$ field direction in those regions, eventhough this field (typically 100 mG) is small enough to avoid broadening
of the transition. We note that those conditions are easier to fulfill than those realized in
\cite{woo97}, since the field direction remains the same between the preparation and the interaction regions (see Fig. 2),
instead of having to be rotated by $\pi/2$. 
\section{Suppression of the systematic effect arising from the $M'_1$ Stark interference via optical birefringences } 
As we may note on Eq. 1, when the  sign of the  true scalar $\vec P_e \cdot \vec E \wedge \hat k$ is reversed, the
discrimination between the 
$E_1^{pv}$-Stark and the $M'_1$-Stark interference signals hinges on their opposite behaviour under
reversal of the pseudoscalar $\xi$, the excitation light helicity. Since in $^{133}$Cs, the latter is the larger of the two signals, by a
factor $2\times 10^4$, this is a major source of potential systematic effect\footnote{From the calculated magnetic dipole
transition amplitudes \cite{joh99}, we can expect $M_1(Fr)/M_1(Cs) \sim 13$,  while from
\cite{fla95} we expect $E_1^{pv}(Fr)/E_1^{pv}(Cs)
\sim 18$ , hence a similar order of magnitude is expected for the ratio
$M_1/E_1^{pv}$ in both alkalis.}. Indeed, APV measurements previously performed in a transverse electric field have all
met the difficulty associated with the  presence of the parity conserving interference effect, which can mimic the PV signal if
the reversal of the light helicity $\xi$ is imperfect, i.e. if a small component of linear polarization changes its sign
simultaneously with $\xi$. This kind of problem occurs when the optics on the path of the excitation beam possess some
birefringence, a defect difficult to avoid completely at the level required.
\begin{figure}
\centerline{\epsfxsize=150mm  \epsfbox{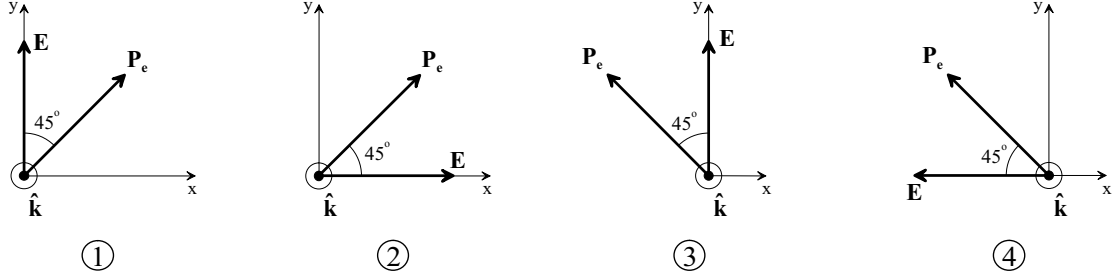}}
\caption{The four geometrical configurations considered in the text, specified by the relative directions of the Stark field,
$\vec E$, the spin polarization created in the ground state, $\vec P_e$, and the wave vector of the excitation 
laser $\hat k$; $\xi $ is assumed the same in the four configurations; the atom velocity is along $\hat k$, orthogonal to the
page. By combining measurements performed in those four configurations, the PV signal is obtained with  considerable
reduction of the systematic effect arising from the
$M'_1$-Stark interference signal via optical birefringences. }
\end{figure}

For the complete discussion given below, we have to write down the expression for the $M'_1$-Stark signal assuming the
most general description of the excitation light polarization. It is expressed in terms of  the four Stokes parameters, which
give a general representation of the beam polarization properties: $u_0 = \vert \epsilon_y \vert^2$ + $\vert
\epsilon_x \vert^2$;  $u_1= {\rm Re} \lbrace \epsilon_x
\epsilon_y^* + \epsilon_x^* \epsilon_y  \rbrace$; $u_2  \equiv \xi = {\rm Im} \lbrace
\epsilon_x^* \epsilon_y - \epsilon_x \epsilon_y^*  \rbrace$; $u_3 = \vert \epsilon_y \vert^2$ -
$\vert \epsilon_x \vert^2$. The first parameter $u_0$ represents the unpolarized intensity. If it is normalized to unity, the
other parameters represent polarization ratios measured by a linear analyzer directed along x, then y  ($u_3$) or along the
bisectors of x and y ($u_1$) or by a direct then inverse circular analyzer ($u_2$).   

According to Eq. 19 of Appendix A, the general expression for the $M'_1$-Stark interference signal ${\cal S}(M_1)$ is given by:
\begin{equation}
{\cal S}(M_1)  =  - 2 M'_1 \, {\rm Re} \lbrace (\beta 
\vec E \wedge \hat k \cdot \hat \epsilon)(\hat \epsilon ^* \cdot \vec P_e) \rbrace \, . 
\end{equation}

We consider the four distinct geometrical configurations represented on Fig.4. Measurements relative to configurations 1
and 2 (or 3 and 4) can be performed simultaneously in the two distinct interaction regions, whereas reversal of $\vec P_e$ by
$\pi/2$ is needed for changing configuration 1 into 3 and 2 into 4.  It is interesting to compare the ${\cal S}(M_1)$ signals
expected in those four configurations:
\begin{eqnarray}
 {\cal S}_1(M_1)= - 2 M'_1 \beta  E \left( \vert \epsilon_x\vert ^2 + {\rm Re}\lbrace{\epsilon_x^* \epsilon_y\rbrace} 
\right)=  -  M'_1 \beta  E \left ( u_0 - u_3 + u_1 \right) \\
{\cal S}_2(M_1)= \;\;\, 2 M'_1 \beta  E \left( \vert \epsilon_y\vert ^2 + {\rm Re}\lbrace{\epsilon_x^* \epsilon_y\rbrace} 
\right)=  \;\;\,  M'_1 \beta  E \left ( u_0 + u_3 + u_1 \right)\\
{\cal S}_3(M_1)= \;\;\, 2 M'_1 \beta  E \left( \vert \epsilon_x\vert ^2 - {\rm Re}\lbrace{\epsilon_x^* \epsilon_y\rbrace} 
\right)=  \;\;\,  M'_1 \beta  E \left ( u_0 - u_3 - u_1 \right) \\
{\cal S}_4(M_1)= -2 M'_1 \beta  E \left( \vert \epsilon_y\vert ^2 - {\rm Re}\lbrace{\epsilon_x^* \epsilon_y\rbrace} 
\right)=  -  M'_1 \beta  E \left ( u_0 +  u_3 - u_1 \right)   
\end{eqnarray}
On the other hand configurations 1 and 2 provide opposite circular dichroism i.e. opposite PV signals, ${\cal S}_1(PV)= -{\cal
S}_2(PV)= \left({\rm Im }E_1^{pv} \beta  \vec E \wedge \xi \hat k \cdot \vec P_e \; \right) $, and the same result holds
for 3 and 4. From the above set of four equations one can form the linear combination  
\begin{equation}
{\cal S}_1  -  {\cal S}_2 -  {\cal S}_3  +  {\cal S}_4 = 4 \left( { \,\cal S}_1(PV) - M'_1 \, \beta  E  \, u_0 \;\right)\,,          
 \end{equation}
which shows up an important property~: the contribution of the $M'_1$-Stark interference signal involves only the unpolarized
intensity,
$u_0$. Thereby when $\xi $ is reversed, so as to isolate the $E_1^{pv}$ contribution, we reduce
considerably the risk which would have come from $\xi-odd$-contributions contaminating either $u_3$ or $u_1$, via the
birefringence of the optics\footnote{More precisely, the birefringence
$\alpha_3$ of axes x and y induces a small polarization $u_1= 2 \alpha_3  \xi$ and the birefringence $\alpha_1$, with
axes oriented at 45$^{\circ}$, a small polarization $u_3 = 2 \alpha_1 \xi$.}. 

 As a convenient and reliable means of performing helicity reversal,  one can use the polarization modulator described in
ref
\cite{bou81}. It provides specific labelings of the three Stokes parameters, $u_1, \; \xi \equiv u_2 $, and $u_3$, by distinct 
modulations. In this way both signals ${\cal S}(PV)$ and  ${\cal S}(M_1)$ appearing at different frequencies are detected by
synchronous detection.     

Additional discrimination of ${\cal S}(PV)$ against  ${\cal S}(M_1)$, respectively even and odd under $\hat k$ reversal, can
be obtained by performing multiple passages of the beam between two mirrors pierced in their center, following a
procedure used in \cite{mul,bou82}. 

\section{Magnitude of the expected signals}
In the preceding sections we have made precise suggestions for adapting APV measurements to a source of cold atoms.
Now, we intend to give an estimate of both the APV and the
$M'_1$-Stark interference signals,
${\cal S}(PV)$ and  ${\cal S}(M_1)$, and their Signal to Noise ratios (SNR), assuming reasonable magnitudes of the Stark field
and the laser intensity. We note that the shot noise limited SNR is independent of the magnitude of the Stark field. 
We take the example of $^{133}$Cs in order to make easier comparison with other atomic sources already exploited.  First, we
need to evaluate the excitation probability per unit of time:
${\cal R}_{ex}=\frac {dN_{Cs}^{*}}{dt}/N_{Cs} =
\sigma_{ex}(E) \times \Phi_{ex} $, where $\Phi_{ex}$ is the flux of excitation photons. The excitation cross section
without electric field,
$\sigma_{nat}=\sigma_{ex}(E=0)$, without Doppler broadening, for an isotope without nuclear spin, excited by a single-mode
laser centered in frequency at the transition peak, is given in ref \cite{bou75}: 
\begin{equation}
 \sigma_{nat} = \frac{ \lambda^2}{2 \pi} \; \frac{\Gamma_{M'_1}}{\Gamma_{7S}} = 2.45
\times 10^{-23}\;  {\rm cm}^2 \,. 
\end{equation}
Here ${\Gamma_{7S}}$ denotes the natural width of the 7S state and $\Gamma_{M'_1}$ the partial width associated
with the $M'_1$ amplitude. Assuming excitation of the $6S_F \rightarrow 7S_{F'} $ line 
in an electric field, using the results of Appendix A, we obtain:
\begin{equation}
 \sigma_{ex}(E)=   
\frac{(2F'+1)}{2(2I+1)} \times \frac{2}{3}(1-g_{_{F'}}) \times \left(\frac{\beta E}{M'_1} \right)^2\sigma_{nat}   
\end{equation}

\subsection{Measurement of $M'_1/\beta E$ with a cold atomic beam (proposal 1) }
 
For a Stark electric field of 1000 V/cm, leading to $\sigma_{ex}(E)= 0.89\times 10^{-20}$ cm$^2$ for the  $6S_{F=3}
\rightarrow 7S_{F'=4}$ line and   $\beta E/M_1= 1000/30$, for an excitation beam of waist radius 1 mm, delivering 500 mW at
539.4~nm ($\Phi_{ex}$ = 0.95
$\times 10^{20}$ photons s$^{-1}$/cm$^2$),  we predict  
${\cal R}_{ex} = 0.89 \times 10^{-20} \times 0.95 \times 10^{20}= 0.84$ s$^{-1}$. Using the number of Cs atoms in the
interaction region, given in Table 1 (proposal 1), we expect $\frac{dN^*_{Cs}}{dt}=0.84 \times 10^7  $s$^{-1}$ for the two
interaction regions, each 20 mm long. 
Supposing 
a fluorescence detection efficiency of 10 $\%$, we predict a collected fluorescence rate of $\sim 10^6 $ s$^{-1}$. Using Eq.
1 (and 2), with
$\vert \vec E \cdot \hat k \wedge \vec P_e \vert = 1/\sqrt{2}$, we expect a SNR$\simeq 15/\sqrt{\rm Hz}$ for ${\cal
S}(M_1)$ for the  $6S_{F=3} \rightarrow 7S_{F'=4}$ line (and $\simeq 20/\sqrt{\rm Hz}$ on the  $6S_{F=4} \rightarrow
7S_{F'=3}$ line). Hence a statistical precision of
$10^{-3}$ can be obtained with an integration time of about one hour.  For the measurement of $M_1^{hf} \sim M'_1/5$
at the same level of precision, the integration time has to be 25 times longer, for both $\Delta F =1$ and $-1$ lines. This
looks possible to achieve. We believe that the conditions for observing this signal could be made excellent~: thanks to the
very good vacuum realized by differential pumping in the beam compartment which is well separated from the MOT by
the pyramidal assembly, we can expect nearly no background. In this respect the signature given to ${\cal
S}(M_1)$ by modulating $u_3$ and $u_1$ (see Eqs 4 to 7) should be of great help.   

On the other hand, with ${\rm Im} E_1^{pv}/\beta E = 1.6 \times 10^{-6}$, there is no chance to achieve APV
measurements without recourse to some amplification process. A possibility might rely on multiple passages of the
excitation beam which can also provide efficient suppression of the $M'_1$-Stark interference signal and hence further 
reduction of the associated systematics. If we denote by
$\kappa$ the signal enhancement factor, the SNR for ${\cal S}(PV)$ becomes  $\sim \sqrt{\kappa} \times 10^{-3}/\sqrt{\rm
Hz}$, hence the time required for observing the PV effect with SNR = 1 is $10^6/\kappa$ seconds. An enhancement factor larger
than 100 would be necessary for obtaining worthwhile conditions of measurement. 

We can now examine the situation with francium. As mentioned earlier we can expect the francium  $M'_1$ amplitude to be
one order of magnitude larger than the cesium one. This increases the ${\cal
S}(M_1)$ without adding noise.  The shot noise limited SNR ratio is thus increased by a
factor of 10. On the other hand, the atom flux will certainly be reduced. The best production rates of Fr$^{+}$ ions available in the
world is, to our knowledge, at the ISOLDE facility at CERN~: it amounts to
$\sim 10 ^9$ s$^{-1}$. We are presently uncertain about the efficiency of neutralization and collection in the MOT, 
$\zeta$, one may expect. A fairly conservative estimate might be $\zeta \sim 10^{-2}$. However, taking into account that a
80$\%$ ion to atom conversion efficiency has been  reported for the converter used on-line at ISOLDE
\cite{tou81} and that a 16$\%$  collection and trapping efficiency has been achieved with Fr atoms \cite{wie96}, we can
reasonably hope that $\zeta
\approx 0.1$ is achievable. The SNR is reduced by $\sqrt\zeta$.  All in all, we can consider that not only does the observation of
the forbidden 7S-8S transition look feasible but so too does a measurement of its magnetic dipole amplitude with an accuracy better
than 10$\%$. This would provide an important test of atomic models \cite{joh99}. Such an experiment would also give invaluable
insight into how to perform a future measurement of $Q_W({\rm Fr})$~: for such a measurement to become possible with   
an efficiency $\zeta=10^{-2}$, the same enhancement factor $\kappa = 100 $ as for cesium is required.
\subsection{ Prospect for APV observation with an  ultra cold atomic beam (proposal 2) }
As shown in Table I, the ultra cold beam can {\it a priori} offer better performances owing to the possibility of lengthening
the interaction time. However, this advantage is spoiled by the effect of the beam divergence, which one would like to
reduce by a factor of $\sim 3$.  One possibility consists in making additional transverse cooling of the
atomic beam simultaneously at the output of the MM-MOT, using an auxiliary 2D MOT according to a
scheme used by the authors of ref\cite{dgo02} for loading the beam into a magnetic guide. If one wants to benefit from 
the lowest velocities, $\sim$ 20 cm/s reported in\cite{dgo02}, {\it a priori} very interesting here, one has to solve the problem
of collisional losses of the slow atomic beam with atoms in the vapor cell on its way to the interaction region, possibly by using 
 other means for loading the trap. 

Another 
important technical question, beyond the scope of the present paper, concerns the possibility of combining the advantages
of multiple passages of the longitudinal excitation beam with those of the ultra-cold atomic beam.

\subsection{APV observation with a Zeeman-slowed atomic beam (proposal 3)}
The number of atoms in the interaction region obtainable with the Zeeman slower is given in table I. It corresponds to a gain
by a factor of $\sqrt{17}$ with respect to the slow and cold atomic beam (proposal 1). The shot noise limited S/N ratio for ${\cal
S}(PV)$, increased by that same factor,  becomes $\sqrt{17 \kappa} \times 10^{-3} /\sqrt{\rm Hz}$. For becoming competitive
with the thermal beam Boulder experiment, from the sole point of view of SNR ratio, an enhancement factor $\kappa$ of $\sim 
6\times 10^3$ is necessary. In this experiment, the collimator causes a  deflection of the atomic beam and a 
Fabry-Perot cavity enhancing the intensity of the excitation  beam all along the interaction region does not
look too unrealistic, but the enhancement factor required for obtaining the same SNR is comparable to that achieved in Boulder.
One may, however, expect that the high power stored inside the cavity will have here somewhat milder drawbacks. Indeed,
longitudinal excitation allows to all the excited atoms to explore the interference pattern over several wavelengths during their
lifetime, hence the difficulty associated with inhomogeneous light shifts causing asymmetric line shapes should be suppressed. In
conclusion, for APV measurements on the stable
$^{133}Cs$ atom the Zeeman slower is an interesting possibility but, with respect to the thermal beam \cite{woo97}, we cannot
expect neither simplifications of the set-up, nor drastic improvement of the SNR ratio.

\section{Conclusion}

In this paper we have addressed the question of how to best use a cold atom source for performing APV measurements. For
fighting against the large drawback associated with the small number of atoms compared with cells, one must take the
maximum advantage of their narrow velocity distribution. This advantage makes it possible to excite a beam of slow and
cold atoms by a (co-)antico-linear laser, spatially matching the atomic beam over several centimeters, without any
Doppler broadening. With respect to a thermal beam, the lengthening of the interaction time thus achieved ranges between
$10^3$ and $10^4$.  On the other hand, we have made a new proposal concerning the observable physical quantity
manifesting APV. The atomic beam should be given a transverse spin polarization,
$\vec P_e$. The new observable reflects existence of a circular dichroism. It involves the pseudoscalar $\vec E\cdot \xi
\hat k \wedge \vec P_e$ and appears in the population of the upper state, hence in the total fluorescence light. Therefore   
fluorescence detection efficiency is a crucial parameter to be optimized. Moreover, with two interaction regions leading to 
opposite circular dichroism, it is possible to make differential measurements. If, in addition, the spin polarization  $\vec
P_e$ can be sequentially rotated by $\pi/2$, then by combining the four results obtained in the two interaction
regions for the two orientations of $\vec P_e$, it is also possible to achieve important reduction of the systematic effects that
birefringence of the optics may generate from the $10^4$ times larger $M_1$-Stark interference signal.
  
The merit of cold atom sources relies on their potential to localize atoms,
only one of the conditions required to
  extend APV measurements in the long term to radioactive isotopes. Suppression of Doppler broadening and lengthening of the
interaction time are other important benefits. However, our estimate of the S/N ratio shows that, in the present state of the
art, those do not appear sufficient 
to solve the difficulty of precise APV measurements. Nevertheless, exploratory experiments
performed on stable alkali highly forbidden transitions, can provide a valuable step enabling us to define the beam specifications
required for APV experiments with radioactive isotopes.   We have shown that by combining experimental techniques
 proven elsewhere, there is a reasonable hope of observing the 6S-7S transition for
$^{133}$Cs and of making a $10^{-3}$ accurate measurement of $M_1^{hf}/\beta$ with a beam of slow, cold atoms with an
unsophisticated set-up. Furthermore, such an experiment could be considered as a prototype to evaluate the production rate of Fr
atoms needed to extend such measurements from stable
$^{133}$Cs to radioactive Fr. With a Zeeman slower providing a monokinetic beam of high flux and low divergence, PV
measurements on
$^{133}$Cs as precise as those presently existing do not look impossible, but a real progress with respect to a thermal beam does
not look obvious to us.  We hope that our present contribution will stimulate
both reflections and experimental work towards advances  in this emerging field of research.

\vspace{5mm}
We thank particularly D. Gu\'ery-Odelin, L. Moi, C. J. Foot, D. Cassettari, A. Camposeo and F. Cervelli for very stimulating
discussions and practical advice for preparing the ongoing construction of a laser trap.
We are grateful to M.D. Plimmer for critical reading of the manuscript. A.W. acknowledges support from CNRS (IN2P3)
and S.S. from the European Commission.

\vspace{5mm}
$^a$ {\footnotesize Also at E. Fermi Physics Dept., Pisa Univ., Pisa, Italy.}

\newpage
 
\section*{Appendix A}  

We are now going to present the derivation of the population signal in the experimental configuration specified in this paper.
The nS, F $\rightarrow $ (n+1)S, F' transition amplitudes can be obtained from the effective transition matrix
$T$ acting on the electronic spin states of the form:
\begin{equation}
T=a 1 \hspace{-0.8mm}{\rm I} + \vec b \cdot \vec \sigma 
\end{equation}
\noindent $1 \hspace{-0.8mm}{\rm I}$ is the two-by-two unit matrix and the components of $\vec \sigma$ are the
three Pauli matrices. The parameters $a$ and $\vec b$ are given by:

\begin{eqnarray}
a & = & \alpha  \vec {\rm E} \cdot \hat \epsilon\\
\vec b & = &  i \beta \; {\rm \vec E} \wedge \hat \epsilon -  M'_1 \hat k \wedge \hat \epsilon + i {\rm Im }
E_1^{pv} \hat \epsilon\;. 
\end{eqnarray}
\noindent  where $\alpha$ and $\beta$ are the scalar and vector transition polarizabilities, $M'_1$ and $E_1^{pv}$ are the
magnetic dipole and the parity-violating electric dipole transition amplitudes, and $\hat \epsilon$ represents the laser
polarization.

In the present situation, stimulated emission is totally negligible compared with spontaneous emission and optical
coherences between the two S states can be ignored. We assume that the laser selects one hfs
component  nS, F $\rightarrow$ (n+1)S, F'. The excited state density matrix, up to a normalization factor, is then given by: 
\begin{equation}
\rho = P_{_{F'}}\,  T P_{_{F}} \,  \rho_g P_{_{F }} \, T^{\dagger} \, P_{_{F'}}\,,
\end{equation}
where  $\rho_g$ is the restriction of the density operator to the nS ground state. $P_{_{F}}$ is the
projector on the nS, F sublevel and $P_{F'}$ the projector on the (n+1)S, F' sublevel. We assume an electronic orientation, 
$\vec P_e$, has been  created in the ground state:
 \begin{equation}
  \rho_g   = 1 \hspace{-0.8mm}{\rm I} + \vec P_e \cdot \vec \sigma\,.
\end{equation}
This definition implies that : ${\rm Tr} \rho_g = 2(2I+1)$.   Therefore, a common normalization 
factor 1/2(2I+1) has to be applied to all the quantities computed below. It is taken into account in Eq. 10. 

In order to compute the 7S population and its spin polarization,
proportional respectively to Tr $\rho$ and Tr $\rho \vec
\sigma$, we apply the Wigner-Eckart theorem to the spin operator $\vec \sigma$ acting in the hyperfine subspace F :  
\begin{equation}
P_{_F} \,  \vec \sigma  P_{ _F } =  2 g_{_F} \, P_{_F} \,  \vec F , \hspace{10mm} {\rm where}  \hspace{5mm} g_{_F} =
2(F-I)/(2I+1). 
\end{equation}

Using Eqs 1 to 6 we obtain for the $\Delta F= F' - F $ transition:
\begin{eqnarray}
&{\rm Tr} \rho = n_{_{F'}} ( \delta_ {_{F F'}} aa^*  + h_{_{F F'}}\,  \vec b\cdot \vec b^*) + [ n_{_{F '}} p_{_{F'}} \delta_{_{F
F'}} (a
\vec b^* + a^ * \vec b \, ) - n_{_{F}} p_{_{F}} g_{_{F' F}} i \vec b \wedge \vec b^*] \cdot \vec P_e\, ,    
\\  &{\rm Tr} \rho \vec \sigma =   n_{_{F '}} p_{_{F'}} [\delta_{_{F
F'}} (a
\vec b^* + a^ * \vec b \, ) + g_{_{F F'}} i\vec b \wedge \vec b^*] +  \vec P_e \; {\rm -dependent \; contributions }\, ,  
\end{eqnarray}
where $ n_{_{F'}} = 2 F' + 1 ,\;  p_{_{F'}}= (1+ 2 g_{_{F'}})/3 , \; $ and, 

if $\Delta F =0, h_{_{F F'}}= p_{_{F'}} $, and 
 $g_{_{F F'}} = g_{_{F'}}$;  

if $\Delta F = \pm 1,  h_{_{F F'}}= 2(1-g_{_{F'}})/3 \equiv 4 g_{_{F}}^2  F(F+1)/3$ and
 $g_{_{F F'}} = 1-g_{_{F'}}$.   

The second term in the RHS of Eq. 17 represents the contribution to the upper state population which depends on the initial
state orientation $\vec P_e$, while Eq. 18 gives the orientation of the upper state created by the excitation process, the
observable physical quantity that we detected in our first APV experiment \cite{bou82}. We note the close connection
between those two contributions in which the role of the initial and final states is interchanged. (Note the appearance of F, and
not F', in the last term of the RHS of Eq. 17). 

Keeping only the terms which depend on the Stark field, we obtain for a $\Delta F=\pm 1$, nS, F $\rightarrow $ (n+1)S, F',
transition:
\begin{eqnarray}
&{\rm Tr} \rho =   
(2F + 1)  \frac{1-g_{_{F}}}{3}  \times  \nonumber \\  & \left[ 2 \beta^2 \vert \vec E \wedge \hat \epsilon \vert^2 
 - (1 + 2 g_{_{F}}) \left({\rm Im }E_1^{pv} \beta  \vec E \wedge \xi \hat k \cdot \vec P_e  + 2 M'_1{\rm
Re}\lbrace (\beta 
\vec E \wedge \hat k \cdot \hat \epsilon)(\hat \epsilon ^* \cdot \vec P_e) \rbrace+ \beta^2 \xi (\vec E \cdot
\hat k ) (\vec E \cdot \vec P_e) \right) \right] \; .
\end{eqnarray}
The last $\vec P_e$-dependent contribution in Eq. 19 vanishes when there is no longitudinal component of the
electric field.

  \end{document}